%% file: main.tex
\documentclass[twocolumn,english,aps,prd,reprint,floatfix,notitlepage,footinbib,preprintnumbers,superscriptaddress,altaffilletter]{revtex4-2}

\usepackage{amsmath,amssymb,amsfonts}
\usepackage{hyperref,breakurl,cleveref,url}
\usepackage{color}

\usepackage{graphicx}
\usepackage[export]{adjustbox}

\usepackage{accents,mathrsfs,mathtools}

\renewcommand{\paragraph}[1]{%
    \textit{#1}.---%
}

\newcommand\trick[1]{}

\usepackage{enumitem}
\setlist[enumerate]{
    label={},
    leftmargin=2em,
    itemsep=2pt,
    topsep= 2pt,
    partopsep=0pt,
    parsep=0pt,
}

\let\oldeqref\eqref
\renewcommand{\eqref}[1]{Eq.\,\smash{\oldeqref{#1}}}
\newcommand{\eqrefs}[2]{Eqs.\,\smash{\oldeqref{#1}}\,and\,\smash{\oldeqref{#2}}}

\newcommand{\rcite}[1]{Ref.\,\cite{#1}}
\newcommand{\rrcite}[1]{Refs.\,\cite{#1}}

\newcommand{\fref}[1]{Fig.\,\ref{#1}}

\def\JHK{{J.-H.\,K. }}

\def\mem{\hspace{0.1em}}
\def\hem{\hspace{0.05em}}
\def\nem{\hspace{-0.1em}}
\def\hnem{\hspace{-0.05em}}
\def\hhem{\hspace{0.025em}}
\def\hhnem{\hspace{-0.025em}}
\def\hhhem{\hspace{0.0125em}}

\def\qiq{{\quad\implies\quad}}

\def\minie{{\textstyle\frac{1}{2}}}

\def\Kerr{{\smash{\text{$\kern-0.075em\sqrt{\text{Kerr\hem}}$}}}}

\def\a{\alpha}
\def\b{\beta}

\def\e{\epsilon}
\def\ve{\varepsilon}
\def\m{\mu}
\def\n{\nu}
\def\r{\rho}
\def\s{\sigma}
\def\k{\kappa}
\def\l{\lambda}

\def\be{{\bar{\epsilon}}}

\def\bpsi{{\smash{\bar{\psi}}\kern0.02em\vphantom{\psi}}}

\def\bz{\bar{z}}
\def\bZ{\bar{Z}}
\def\mathe{{\scalebox{1.03}[1]{$\mathrm{e}$}}}

\def\mdot{{\mem\cdot\mem}}
\def\mwedge{{\mem\wedge\mem\hhem}}

\def\da{{\dot{\a}}}
\def\db{{\dot{\b}}}

\def\rmA{{\mathrm{A}}}
\def\rmB{{\mathrm{B}}}

\newcommand{\wrap}[1]{{\smash{#1}\vphantom{\beta}}}

\newcommand{\Ket}[1]{{\hem\big|\hem{#1}\big\rangle}}
\newcommand{\Bra}[1]{{\big\langle{#1}\hem\big|\hem}}
\newcommand{\BraKet}[2]{{\big\langle{#1}\hem\big|\hem{#2}\big\rangle}}

\def\rmx{{
    \scalebox{1.2}[1]{$\mathrm{x}$}\kern-0.625em\scalebox{1.2}[1]{$\mathrm{x}$}
}}
\def\rmS{{
    \scalebox{1.09}[0.95]{$\mathrm{S}$}\kern-0.625em\scalebox{1.09}[0.95]{$\mathrm{S}$}
}}
\def\rmN{{
    \scalebox{1.2}[0.95]{$\mathrm{N}$}\kern-0.895em\scalebox{1.2}[0.95]{$\mathrm{N}$}
}}

\def\P{{\mathcal{P}\kern-0.18em}}

\let\del\undefined

\newcommand{\del}[2]{
    \delta^{(4)}\hnem\big(\hem\hhhem{
        {#1}
        {\mem-\mem}
        {#2}
    }\hem\big)
}
\newcommand{\dell}[2]{
    \delta^{(4)}\hnem\big(\hem\hhhem{
        {#1}
        - 
        {#2}
    }\hem\big)
}

\newcommand{\inv}[1]{\mathrlap{\smash{\adjustbox{raise=0.11em}{$^{\mem-1}$}}}_{\mathrlap{#1}\hphantom{\mem-1}}\hnem}

\def\rambda{{\bar{\lambda}}}
\def\bmu{{\bar{\mu}}}

\begin{document}

\title{
    Asymptotic Spinspacetime
}

\author{Joon-Hwi Kim}
\affiliation{Walter Burke Institute for Theoretical Physics, California Institute of Technology, Pasadena, CA 91125}

\begin{abstract}
    We show that 
    Poincar\'e invariance directly implies the existence of
    a complexified Minkowski space whose real and imaginary directions unify spacetime and spin, which we dub spinspacetime.
    Despite the intrinsic noncommutativity of spin,
    spinspacetime exhibits
    mutually commuting
    holomorphic coordinates.
    Its twistorial construction
    derives the Newman-Janis shift property of spinning black holes 
    by massive half-Fourier transforming
    complexified on-shell kinematics,
    which encode a spinning analog of equivalence principle.
\end{abstract}

\preprint{CALT-TH 2023-038}


\bibliographystyle{utphys-modified}

\renewcommand*{\bibfont}{\fontsize{7}{7.5}\selectfont}
\setlength{\bibsep}{1pt}

\maketitle

\paragraph{Introduction}%
In the 70s,
Ezra Ted Newman
investigated
the curious idea of
unifying spacetime and spin into a complex geometry
\cite{newman1974curiosity,newman1974collection,newman1973complex,newman2004maxwell,Newman:1973yu,Newman:2002mk,ko1981theory,grg207flaherty}.
While offering
a unique perspective
on relativistic angular momenta,
it 
provided
a suggestive 
argument for
the Newman-Janis \cite{Newman:1965tw-janis,Newman:1965my-kerrmetric} derivation of spinning black hole solutions
where spin turns into an imaginary deviation
\cite{newman1974curiosity,newman1974collection,newman1973complex,newman2004maxwell,Newman:1973yu,Newman:2002mk,ko1981theory,grg207flaherty}.
However, it remains uncertain if
the Newman-Janis shift really follows solely from this complex geometry,
since Newman's argument implicitly counts on an assumption
that the black hole would be
localized in the complex 
for an unknown reason.
On a related note,
it has been also not clearly understood
whether this concept is necessarily limited to black holes or not.
Unfortunately,
this line of ideas
appears to be
no longer
actively revived in the current literature.

In this paper,
we aim to give a dedicated spotlight on this ``forgotten lore''
and unveil its relevance to
the modern program of
describing spinning black holes
in terms of scattering amplitudes
\cite{ahh2017,aho2020,Guevara:2018wpp,Guevara:2019fsj,chkl2019,Johansson:2019dnu,Lazopoulos:2021mna,Aoude:2020onz,aoude2022searching,Cangemi:2022bew,Cangemi:2023bpe,gmoov,ambikerr1,Kim:2024grz}.
As the very first step towards this modern resurrection,
we endow it with a name: ``spinspacetime.''
We provide
a complete definition of spinspacetime
that unifies various perspectives
\cite{newman1974curiosity,newman1974collection,newman1973complex,newman2004maxwell,Newman:1973yu,Newman:2002mk,ko1981theory,grg207flaherty,almond1973time,casalbuoni1976classical,hughston1979twistors,bette1984pointlike,bette1996directly,bette1997extended,bette2000twistor,bette2004massive,bette2004massive04,bette2004massive05,bette2005twistors-0402150,bette2005two-0503134,lukierski2014noncommutative,Filippas:2022soduality}.
Especially, we highlight its characteristic commutator structure,
which, to our knowledge, was not investigated in Newman's works
\cite{newman1974curiosity,newman1974collection,newman1973complex,newman2004maxwell,Newman:1973yu,Newman:2002mk,ko1981theory,grg207flaherty}.
Consequently,
we refine Newman's ideas
to precise statements.
In particular,
we present
a new derivation of the Newman-Janis shift from scattering amplitudes,
which may trigger
an extension of the on-shell/twistor diagrams
\cite{hodges1980twistor,hodges1990feynman,hodges2005a,hodges2005b,arkani2010s-matrix,arkani2012positivegrassmannian,Atiyah:2017erd}
to massive states.

Our central findings are twofold.
Firstly, we show that spinspacetime \textit{universally} arises in
any massive system
exhibiting global Poincar\'e symmetry:
particles and fields in asymptotically flat spacetimes,
or even systems that may lack a spacetime formulation.
We construct real variables $x^\m$, $y^\m$
directly from the Poincar\'e charges.
Then
Poincar\'e symmetry dictates
their commutators to be
\begin{align}
\begin{split}
    \label{eq:prop.zigzag}
    [ z , z ] = 0
    \,,\quad
    [ z , \bz ] \neq 0
    \,,\quad
    [ \bz , \bz ] = 0
    \,,
\end{split}
\end{align}
if $z^\m = x^\m + iy^\m$
and $\bz^\m = x^\m - iy^\m$.
These complex coordinates unify spacetime and spin as real and imaginary parts,
which can be seen from the following split of the total angular momentum into orbital and spin parts:
\begin{align}
    \label{eq:prop.J}
    J = (x\wedge p) + *(y\wedge p)
    \qiq
    J^+ = (z\wedge p)^+
    \,.
\end{align}
Here,
an intrinsic association between chirality and holomorphy
is implied,
as the self-dual part of the total angular momentum, $J^+$, arises solely from the holomorphic coordinates $z$
as $* \to +i$.
Therefore, a complexified Minkowski space 
exhibiting a fascinating geometry
arises,
whose complex coordinates $z^\m$, $\bz^\m$
(1) realize a particular ``off-diagonal'' form of commutators as in \eqref{eq:prop.zigzag}
and (2) are inherently linked with chirality
as in \eqref{eq:prop.J}.
This provides the complete characterization of spinspacetime.

\begin{figure}[t]
    \centering
    \begin{align*}
        \includegraphics[valign=c,scale=1.0]{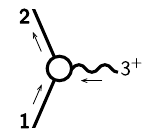}
        &
        \adjustbox{scale=1.1}{$
            \quad\sim\quad
            \del{
                \lambda_1
            }{
                \lambda_2
            }
        $}
        \\
        \includegraphics[valign=c,scale=1.0]{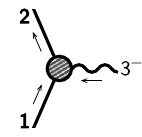}
        &
        \adjustbox{scale=1.1}{$
            \quad\sim\quad
            \del{
                \rambda_1
            }{
                \rambda_2
            }
        $}
    \end{align*}
    \caption{
        The building blocks of 
        ``massive on-shell diagrams,''
        describing extremal cases of complexified on-shell kinematics.
    }
    \label{fig:onshell}
\end{figure}

Secondly,
we specialize this framework
in the scattering problem of massive spinning states
from asymptotically free regions.
To this end,
we adopt a twistorial implementation of spinspacetime,
in which case the off-diagonal form of \eqref{eq:prop.zigzag}
originates from the oscillator commutator algebra
(K\"ahler geometry)
\cite{penrose1968twistor,penrose1973twistor,penrose1975aims}
of twistor space:
\begin{align}
    \label{eq:ZZpb}
    [ Z , Z ] = 0
    \,,\quad
    [ Z , \bZ ] \neq 0
    \,,\quad
    [ \bZ , \bZ ] = 0
    \,.
\end{align}
In this description,
the Newman-Janis shift
(as localizations of scattering amplitudes at complex positions
\cite{ahh2017,aho2020,Guevara:2018wpp,Guevara:2019fsj,chkl2019,Johansson:2019dnu,Lazopoulos:2021mna,Aoude:2020onz,aoude2022searching})
is directly derived from
complexified on-shell kinematics
in \fref{fig:onshell}
via \textit{massive} half-Fourier transforms.
As they
provide the massive counterpart of the MHV kinematics
\cite{arkani2012positivegrassmannian}
and describe a spinning analog of equivalence principle,
a novel perspective
is provided for
the minimality 
\cite{ahh2017,aho2020,Guevara:2018wpp,Guevara:2019fsj,chkl2019}
of spinning black holes.

In conclusion, spinspacetime is a universal framework 
for any massive Poincar\'e-invariant system, 
in which the minimality of spinning black holes is strikingly manifest through its distinctive complex-geometrical structures.
This refined understanding fully resolves the questions left 
by Newman's works
and allows us to extend and generalize its curious story
with exciting future applications.


\paragraph{Reconstructing Spacetime from Poincar\'e Symmetry}%
In any system enjoying global Poincar\'e symmetry,
there shall exist Poincar\'e charges:
\begin{align}
\begin{split}
    \label{eq:pa}
    [ J^{\m\n} , J^{\r\s} ]
    &= i\mem (-4 \delta^{[\m}{}_{[\k} \eta^{\n][\r} \delta^{\s]}{}_{\l]})\hem J^{\k\l}
    \,,\\
    [ J^{\m\n} , p_\r ]
    &= i\mem p_\s\hem (-2\eta^{\s[\m} \delta^{\n]}{}_\r)
    \,,\quad
    [ p_\m , p_\n ]
    = 0
    \vphantom{(-4 \delta^{[\m}{}_{[\k} \eta^{\n][\r} \delta^{\s]}{}_{\l]})\hem J^{\k\l}}
    \,.
\end{split}
\end{align}
Additionally,
the notion of mass dimension is accompanied
in any physical situation.
This amounts to having an operator $D$ such that
\begin{align}
    \label{eq:D|Jp}
    [ J^{\m\n} , D ] = 0
    \,,\quad
    [ p_\m , D ] = +i\mem p_\m
    \,,
\end{align}
assigning dimensions $0$ and $+1$ to $J$ and $p$, respectively.

Suppose the system is massive so that
its configurations 
do not realize $p^2 = 0$.
Then, define
\begin{align}
    \label{eq:xdef}
    x^\m
    = 
    J^{\m\n} p_\n /p^2
    -
    D\hem p^\m \nem/p^2
    \,,
\end{align}
as a function of $J$, $p$, and $D$.
By straightforward calculations,
it follows that
$x^\m$ exactly transforms like a position vector in Minkowski space
under the Poincar\'e group action:
$[ x^\m , J^{\r\s} ] = 
    i\mem
    (-2\eta^{\m[\r} \delta^{\s]}{}_\n)\mem
    x^\n
$,
$[ x^\m , p_\r ] = 
    i\mem
    \delta^\m{}_\r
$.
In this manner,
a Minkowski space
emerges
from Poincar\'e symmetry.

It should be clear that this construction applies to 
\textit{any} massive system with global Poincar\'e symmetry:
a particle, a group of particles, 
a field, a group of fields,
or even a system not prescribed in a spacetime formulation.
The prime example is
a massive system in an asymptotically flat spacetime
\footnote{
    The supertranslation ambiguity of angular momentum
    may be a topic to consider
    in this context
    \cite{Bondi:1962px,Sachs:1962wk,seminardg,Hawking:2016sgy,Szabados:2004xxa}.
},
in which case
our argument asserts that 
a \textit{flat bulk}
emerges
from the Poincar\'e charges at asymptotic infinity.
%

A few remarks are in order.
First, suppose we have only assumed the strict Poincar\'e algebra in \eqref{eq:pa}.
Then $J^{\m\n}p_\n/p^2$ behaves almost, but not perfectly, like a position vector.
Amusingly,
taking \eqref{eq:xdef} then as an ansatz,
imposing the desired transformation behavior of $x^\m$
mandates the commutation relations in \eqref{eq:D|Jp},
so the same construction is reproduced.
%
%
Next,
at the technical level,
the above calculations
will closely parallel
those carried out in \rrcite{brooke1985relativistic,bacry1967space}.
However,
\rrcite{brooke1985relativistic,bacry1967space}
specifically restricted
their scope
to a free particle,
whereas
our very point here is the universality of this construction.

\paragraph{Spin-induced Spacetime Fuzziness}%
Next, we observe
a puzzling feature
in this notion of an emergent flat spacetime:
it is noncommutative as
\begin{align}
    \label{eq:[xx]}
    [x^\m , x^\n]
    = -\frac{i}{p^2}\mem S^{\m\n}
    \,,
\end{align}
which directly follows from 
Eqs.\,\oldeqref{eq:pa}-\oldeqref{eq:xdef}.
Here
$S^{\m\n}$
denotes the transverse projection of $J^{\m\n}$
such that
$p_\m S^{\m\n} = 0$.
Namely, it is
the spin angular momentum
which by definition is
the part of
the Lorentz charge
$J^{\m\n}$
unexplainable by the orbital angular momentum 
computed with spacetime coordinates:
$J^{\m\n} = 2x^{[\m}p^{\n]} + S^{\m\n}$.

This peculiarity is inevitable in the sense that
\eqref{eq:xdef} is the only Poincar\'e-covariant and real spacetime coordinates
that can be directly constructed from $J^{\m\n}$, $p_\m$, and $D$,
as implied by the analysis provided in the supplemental material.
Indeed, the noncommutativity in \eqref{eq:[xx]}
has been observed in
a considerable amount of work
\cite{Born:1935ap,pryce1948mass,fleming1965covariant,Hanson:1974qy,Casalbuoni:1975hx,casalbuoni1976classical,Barut:1980mv,brooke1985relativistic,zakrzewski1995extended,GuzmanRamirez:2013ynp,ramirez2015lagrangian,Deriglazov:2017jub,Deriglazov:2019vcj,hughston1979twistors,bette1984pointlike,bette1996directly,bette1997extended,bette2000twistor,bette2004massive,bette2004massive04,bette2004massive05,bette2005twistors-0402150,bette2005two-0503134,lukierski2014noncommutative,Filippas:2022soduality}
from various angles
and
from numerous spinning particle models.

To precisely understand the issue in a concrete setup,
let us consider a specific Hamiltonian formulation of
a single massive spinning particle
in special relativity
\cite{Hanson:1974qy},
the traditional formulation of which involves
three gauge redundancies related to spin
\cite{Hanson:1974qy,steinhoff2015spin,ambikerr0}.
Although the spacetime coordinates in the physical phase space
are noncommutative (nonzero Dirac brackets \cite{Hanson:1974qy}),
there do exist two widely-adopted resolutions.
Firstly, one \textit{introduces gauge redundancies}
in the description
by working in the unconstrained phase space,
which offers Poincar\'e-covariant and commutative spacetime coordinates.
Hence, in the quantize-then-constrain approach,
one achieves $[x^\m,x^\n] {\:=\:} 0$
so that the familiar notion of position eigenstates ``$\hem|x\rangle$''
or position-represented wavefunctions ``$\psi(x)$''
arise.
Upon second quantization,
this also retrieves the standard quantum field theory setup
where $x^\m$ are labels instead of operators.
The gauge constraints
on physical wavefunction
translates to a field equation 
(cf. \rcite{casalbuoni1976classical}).
Secondly, one \textit{fixes the gauge by a noncovariant condition}.
This harms manifest Poincar\'e covariance
by introducing a reference vector
as explicated in the supplemental material
(Pryce-Newton-Wigner spin supplementary condition \cite{pryce1948mass,Newton:1949cq,steinhoff2015spin}).

In our view, none of these solutions are fundamental
in light of manifest 
gauge and Poincar\'e invariances.
Instead,
we find that
this tension 
(cf.\,\cite{arkani2018locality,Trnka:2013the})
between 
commutativity, 
Poincar\'e invariance,
and gauge invariance
is resolved
by ascending to a new type of geometry.

\paragraph{Unification of Spacetime and Spin}%
\label{sec:spinspacetime}%
In the co-moving frame of the momentum,
\eqref{eq:[xx]} 
boils down to
$[x^i,x^j] = (i\hbar / m^2c^2)\mem \ve^{ij}{}_k\mem S^k$,
which implies a noncommutativity scale of
$\mathit{\Delta}x \sim s^{1/2}\mem (\hbar/mc)$ for spin $s$.
Here, we have 
temporarily
restored 
the fundamental constants.
%
Meanwhile, recall the famous commutator,
$[S^i,S^j] = i\hbar\mem \ve^{ij}{}_k\mem S^k$.
Amusingly, these right-hand sides 
can be canceled by 
composing complex combinations $x^i \mp i\mem S^i\nem/mc$,
precisely because the imaginary unit squares to $-1$.

Let us implement this solution to the noncommutativity puzzle
in a fully Poincar\'e-covariant way.
Define the spin length pseudovector,
which describes
the Pauli-Lubanski psueodvector normalized in the units of length:
\begin{align}
    \label{eq:ydef}
    y^\m = - {*J}^{\m\n} p_\n /p^2
    \qiq
    S^{\m\n} = \ve^{\m\n\r\s} y_r\hhem p_\s
    \,.
\end{align}
Then \eqref{eq:pa} implies
that $y^\m$ transforms like a tangent vector:
$
    [ y^\m , J^{\r\s} ]
    = 
    i\mem
    (-2\eta^{\m[\r} \delta^{\s]}{}_\n)\mem
    y^\n
$, $
    [ y^\m , p_\r ]
    =
    0
$.
In turn,
for $z^\m = x^\m + iy^\m$,
\eqrefs{eq:pa}{eq:D|Jp} imply
\begin{align}
\begin{split}
    \label{eq:zzpb}
    [ z^\m , z^\n ]
    &= [ \bz^\m ,\bz^\n ]
    = 0
    \,,\\
    [ z^\m , \bz^\n ]
    &
    = 
    -\frac{2}{p^2}\mem
    \Big(\mem{
        2y^{(\m} p^{\n)}
        + i\mem \ve^{\m\n\r\s}\hhnem y_\r\hhem p_\s
    }\mem\Big)
    \,.
\end{split}
\end{align}
Remarkably, $z^\m$ themselves are
\textit{commutative}.

In conclusion,
a complexified Minkowski space
\cite{penrose1967twistoralgebra}
\footnote{
    Mathematically, this is the space $\mathbb{C}^4$
    equipped with the flat metric $\eta^\mathbb{C}$ whose signature reduces to Lorentzian on the real section $z^\m {\,=\,}\protect\bz^\m$
    \cite{Adamo2017lectures}.
    One may regard it as the tangent bundle of Minkowski space
    equipped with the adapted complex structure\cite{guillemin1991grauert,guillemin1992grauert,lempert1991global,szHoke1991complex,hall2011adapted}.
},
equipped with 
the commutators in \eqref{eq:zzpb}
featuring
commutative holomorphic coordinates,
can be constructed in any massive system with global Poincar\'e symmetry.
This complexified Minkowski space is the \textit{spinspacetime}.

One might wonder if the use of the imaginary unit,
seemingly a bit radical,
is really necessary.
However, as we have shown in a bootstrap analysis in
the supplemental material,
this is \textit{the only Poincar\'e-covariant solution} to the noncommutativity puzzle.
This uniqueness seems to have not been realized explicitly,
although proposals for the complex-valued coordinates has existed
\cite{almond1973time,casalbuoni1976classical}.

Thus,
it appears that Poincar\'e invariance intrinsically entails a curious complex geometry,
which we then must pay attention to.
Especially,
it is easily checked that
the Poincar\'e group action preserves the complex structure:
it does not mix up holomorphic and anti-holomorphic variables.
Hence there is a chance for
this complex structure, or holomorphy, of spinspacetime 
acquiring a physical significance
as a frame-independent notion.

\paragraph{Newman's Derivation of Spinspacetime}%
Indeed,
holomorphy in spinspacetime
is intrinsically associated with
\textit{chirality}
\footnote{
    The easiest way of seeing this is to realize that
    $y^\m$ is a pseudovector:
    parity flip
    induces complex conjugation, $z^\m {\,\leftrightarrow\,} \protect\bz^\m$.
}.
This point is emphasized in Newman's original derivation of spinspacetime \cite{newman1974curiosity},
sketched in \eqref{eq:prop.J}.
To recapitulate,
Newman 
derives the complex coordinates by
observing
that
the self-dual and anti-self-dual parts of 
the angular momentum
are
$J^\pm = ((x{\,\pm\,}iy) \mwedge p)^\pm$,
where 
$(\a \mwedge \b)^\pm$ denotes the self-dual/anti-self-dual projection of the bivector $(\a \mwedge \b)^{\m\n} = \a^\m \b^\n - \a^\n \b^\m$.
Crucially, 
the self-dual part $J^+$ depends solely on the holomorphic position,
while
the \textit{anti}-self-dual part $J^-$ depends solely on the \textit{anti}-holomorphic position.
In this light,
the imaginary unit in $x {\,\pm\,} iy$
has originated from
a Hodge (electric-magnetic \cite{Filippas:2022soduality}) duality between
orbital and spin angular momenta,
$(x \mwedge p)$
and
${*}(y \mwedge p)$.

Newman \cite{newman1974curiosity}'s approach to spinspacetime
is rather independent from 
the approach in the previous sections
where the focus is put on the commutator structure
\cite{almond1973time,casalbuoni1976classical}.
However,
we emphasize that
both perspectives capture essential defining features of spinspacetime:
the geometry of spinspacetime
is completely characterized only if \textit{both}
(1)~%
the commutator structure in \eqref{eq:prop.zigzag}
(more precisely \eqref{eq:zzpb})
and
(2)~%
the association between holomorphy and chirality
tracing back to \eqref{eq:prop.J}
are understood.

These two unique geometrical features 
lead to remarkable physical implications.
They can be demonstrated nicely in a twistorial implementation of spinspacetime,
which we now elaborate on.

\paragraph{Spinspacetime in Massive Twistor Space}%
The astute reader
will realize that
the 
development
so far strongly resonates with the very spirit of twistor theory.
First of all,
spacetime is regarded as a secondary construct
\cite{penrose1987origins,penrose1975aims,penrose1973twistor}.
And not to mention,
a complex-geometrical structure is disclosed.
Further,
the $[z^\m,\bz^\n]$ bracket
takes a simple form in the spinor notation,
as will be revealed shortly in \eqref{eq:z|bz}.
In addition, $D = -p_\m\hem x^\m$ in \eqref{eq:D|Jp} and\:(\ref{eq:xdef})
is the dilatation charge.
These all point to an intimate connection to twistor theory
where complex geometry, spinors, and conformal algebra
play central roles.

A definition of twistor space is
the vector space $\mathbb{C}^4$
equipped with a $(2,2)$-signature Hermitian form,
serving as
the linear representation space of $\mathrm{U}(2,2)$
\cite{penrose1973twistor,penrose1975aims}.
Geometrically,
a twistor represents a null ray in complexified Minkowski space \cite{penrose1973twistor,penrose1975aims,penrose1967twistoralgebra}.
To implement the mass,
one considers
two copies of twistor space,
$\mathbb{C}^8 \cong \mathbb{C}^4 {\mem\times\mem} \mathbb{C}^4$
\cite{Penrose:1974di,Perjes:1974ra,tod1976two},
where $\mathrm{U}(2,2)$
acts from the left
and $\mathrm{U}(2)$
acts from the right
with a shared $\mathrm{U}(1)$.
In essence, this realizes 
a massive momentum as a sum of two null momenta.
This space $\mathbb{C}^8$ is the \textit{massive twistor space}.
According to
a modern view 
developed in
\rrcite{ambikerr0,ambikerr1},
which differs a bit from the original interpretation taken in
the 70's twistor particle program
\cite{Perjes:1974ra,Penrose:1974di},
the right $\mathrm{SU}(2)$ is the massive little group
while the shared $\mathrm{U}(1)$ is gauge.

Having defined the massive twistor space,
the next step is to
construct its spinspacetime coordinates explicitly.
Let $Z_\rmA{}^I = (\lambda_\a{}^I, i\mu^{\da I})$
and $\bZ_I{}^\rmA = (-i\bmu_I{}^\a , \rambda_{I\da})$
be holomorphic and anti-holomorphic coordinates
of the massive twistor space.
Here
$\rmA,\rmB,\cdots$ 
are $\mathrm{SU}(2,2)$ (Dirac spinor) indices,
while
$I,J,\cdots$
are $\mathrm{SU}(2)$ indices.
This space is a K\"ahler vector space
\cite{penrose1968twistor,penrose1973twistor,penrose1975aims},
so
the commutation relations are given by the oscillator algebra
as sketched in \eqref{eq:ZZpb}:
$
    [ Z_\rmA{}^I , \bZ_J{}^\rmB ]
    = \delta_\rmA{}^\rmB\mem \delta_J{}^I
$,
to be explicit.

The $\mathrm{U}(2,2)$ generators
are given by 
$G_\rmA{}^\rmB = Z_\rmA{}^I \bZ_I{}^\rmA $,
the Weyl block decomposition of which gives
\begin{align}
\begin{split}
    \label{eq:pG-twistor}
    p_{\a\da}
    = -\lambda_\a{}^I \rambda_{I\da}
    \,,\quad
    G^\da{}_\db
    = \mu^{\da I} \rambda_{I\db}
    \,.
\end{split}
\end{align}
The first equation is the famous
``massive spinor-helicity'' decomposition
of the massive momentum
\cite{Penrose:1974di,Perjes:1974ra,ahh2017,conde2016spinor,conde2016lorentz}.
The second equation 
unifies dilatation and self-dual angular momentum:
\smash{$
    \smash{G^\da{}_\db}
    = \smash{J^\da{}_\db}
    + \minie\mem \smash{\delta^\da{}_\db}\mem 
    (D {\,+\,} i\tilde{D})
$}.
A new ingredient is $\tilde{D}$,
which generates the $\mathrm{U}(1)$ gauge group.
It has zero commutator with any other generators
and induces
a central extension from $\mathrm{SU}(2,2)$ to $\mathrm{U}(2,2)$.

With this understanding,
we translate
our earlier formula for
spinspacetime coordinates
into the spinor notation.
Notably, it
takes a strikingly simple form:
\begin{align}
    \label{eq:z=Gp}
    z^{\da\a} = -G^\da{}_\db\hem (p^{-1})^{\db\a}
    \,.
\end{align}
Here,
\smash{$(p^{-1})^{\da\a} = \smash{\be^{\da\db} \e^{\a\b}} p_{\wrap{\b\db}} / \det(p)$}
denotes the inverse of $p_{\a\da}$ as a square matrix,
defined for $p^2 \neq 0$.
In turn,
the conformal algebra
directly implies that
\begin{align}
\begin{split}
    \label{eq:z|bz}
    [ z^{\da\a} , z^{\db\b} ] 
    = 0
    \,,\quad
    [ z^{\da\a} , \bz^{\db\b} ]
    &= i\mem (z {\,-\,} \bz)^{\da\b} (p^{-1})^{\db\a}
    \,,
\end{split}
\end{align}
reproducing \eqref{eq:zzpb}
up to the extension $y^\m \mapsto y^\m - \tilde{D}\mem p^\m \nem/p^2$
\footnote{
    \eqrefs{eq:z=Gp}{eq:z|bz} apply
    generally to any
    Hamiltonian system in asymptotically flat spacetime,
    where
    one can find conformal generators
    from the conformal killing vectors
    (even though the 
    Hamiltonian
    may not enjoy
    the full conformal symmetry).
}.
Finally,
applying \eqref{eq:z=Gp}
to
the charges in
\eqref{eq:pG-twistor},
the spinspacetime coordinates 
for the massive twistor space
are found as
\begin{align}
    \label{eq:incidence}
    z^{\da\a}
    = \mu^{\da I} (\lambda^{-1})_I{}^\a
    \qiq
    \mu^{\da I} = z^{\da\a} \lambda_\a{}^I
    \,.
\end{align}
Interestingly,
\eqref{eq:incidence}
reinvents the incidence relation of twistor theory
\cite{penrose1973twistor,penrose1967twistoralgebra,newman1974curiosity}
for the massive case.
Its interpretation
is that
the two complexified null rays represented by
the twistors
$Z_\rmA{}^{I=0,1}$
are ``co-incident''
at a point $z^{\da\a}$
in complexified Minkowski space.

\begin{figure}[t]
    \centering
    \includegraphics[width=0.85\linewidth,trim=0 20pt 0 10pt]{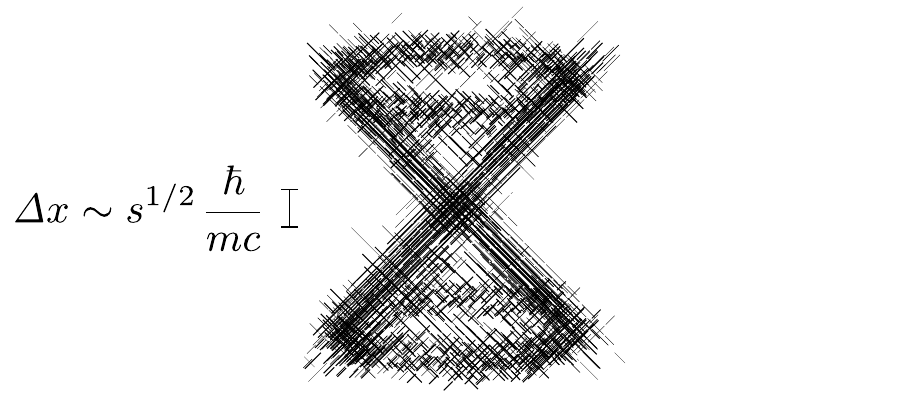}
    \caption{
        Lightcone in twistor theory \cite{penrose1975aims,penrose2005road}.
        Penrose advocates a point of view that
        there is an absurdity inherent in the notion of spacetime
        already at the Compton scale of elementary particles \cite{Penrose:1974di,Penrose:1980hi,penrose1987origins}.
        Accordingly,
        the very philosophy of
        twistor theory
        is to
        take 
        spacetime points as
        secondary constructs
        emerging from intersections between null rays.
    }
    \label{fig:LCdx}
\end{figure}

In conclusion,
the spinspacetime of massive twistor space
is simply the complexified Minkowski space 
which its incidence relation in \eqref{eq:incidence} defines
\footnote{
    Note that a double fibration picture could be pursued to formulate this relationship in a more mathematically precise sense,
    analogously to the massless case.
}.
Intriguingly, this
implies that
the noncommutativity in \eqref{eq:[xx]}
could be reinterpreted
as the fuzziness which spacetime points acquires in twistor theory
as light rays become fundamental objects
\cite{penrose1987origins,penrose1975aims,penrose1973twistor,hughston1979twistors,bette1984pointlike,bette1996directly,bette1997extended,bette2000twistor,bette2004massive,bette2004massive04,bette2004massive05,bette2005twistors-0402150,bette2005two-0503134,lukierski2014noncommutative}:
see Fig.\,\ref{fig:LCdx}.

The twistorial picture
associates fascinating geometric origins to the defining features of spinspacetime.
First,
the off-diagonal commutators
(\eqref{eq:prop.zigzag})
has signaled
the K\"ahler geometry of twistor space
(\eqref{eq:ZZpb}).
Second,
holomorphy in spinspacetime
corresponds to
the holomorphy of twistors.
Third,
the link between holomorphy and chirality in spinspacetime
(\eqref{eq:prop.J})
traces back to
that of
twistor space.
Note that this link is deeply endemic in twistor space:
the \textit{holomorphic} $Z_\rmA{}^I$
contains the \textit{left-handed} spinor-helicity variable $\lambda_\a{}^I$.

\paragraph{S-Matrix in Massive Twistor Space}%
\label{sec:fourier}%
Eventually,
we study
scattering in asymptotically flat spacetimes.
This setup concerns
the spinspacetime 
which
Poincar\'e charges 
in asymptotically free regions
reconstruct,
together with
an asymptotic massive twistor space as a \textit{phase space}.

As is well-known,
a massive on-shell scattering state
is represented by
the massive spinor-helicity variables
$\lambda_\a{}^I$, $\rambda_{I\da}$
\cite{ahh2017}.
The massive twistor space
can be viewed as the cotangent bundle
of the space $\mathbb{C}^4$ of massive spinor-helicity variables,
in light of the commutators
\smash{$
    [ \lambda_\a{}^I , \bmu_J{}^\b ]
    = i\mem \delta_\a{}^\b\mem \delta_J{}^I
$}
and
\smash{$
    [ \rambda_{I\da} , \mu^{\db J} ]
    = i\mem \delta_I{}^J\mem \delta^\db{}_\da
$}.

The massive twistor space admits various polarizations,
offering
different bases for the scattering states.
For instance,
the scattering states can be realized in the coherent state basis
(K\"ahler polarization),
\begin{align}
\begin{split}
    Z \Ket{Z_1}
    = \Ket{Z_1}\mem Z_1
    \,,\quad
    \Bra{\bZ_2} \bZ
    = \bZ_2\hem\hhem \Bra{\bZ_2} 
    \,,
\end{split}
\end{align}
or in the spinor-helicity basis,
\begin{align}
    \lambda \mem\Ket{\lambda_1\rambda_1}
    = \Ket{\lambda_1\rambda_1}\mem \lambda_1
    \,,\quad
    \rambda \mem\Ket{\lambda_1\rambda_1}
    = \Ket{\lambda_1\rambda_1}\mem \rambda_1
    \,,
\end{align}
where we have started to omit indices to avoid clutter.
The overlaps are given by
$\langle \bZ_2 | Z_1 \rangle = \smash{e^{\bZ_2 Z_1}}$
and
$\langle \lambda_2\rambda_2 | \lambda_1\rambda_1 \rangle = \delta^{(4)}({\lambda_1}{\mem-\,}{\lambda_2}) \, \delta^{(4)}({\rambda_1}{\mem-\,}{\rambda_2})$,
where $\bZ_2 Z_1$ abbreviates $(\bZ_2)_I{}^\rmA (Z_1)_\rmA{}^I$.
The transformations between these two bases
are given by the half-Fourier transforms,
which one may define concretely
by analytically continuing to the $(2,2)$-signature
\cite{witten2004perturbative,arkani2010s-matrix}:
\begin{align}
\begin{split}
    \label{eq:half-Fourier}
    \Ket{Z_1}
    = \Ket{\lambda_1\mu_1}
    &= \int\, [d^4\rambda]\,\,
        \mathe^{i\rambda\mu_1}
        \,\, \Ket{\lambda_1\rambda}
    \,,\\
    \Bra{\bZ_2}
    = \Bra{\rambda_2\bmu_2}
    &= \int\, [d^4\lambda]\,\,
        \mathe^{-i\bmu_2\lambda\nem}
        \,\, \Bra{\lambda\rambda_2}
    \,.
\end{split}
\end{align}
In turn,
S-matrix elements in the spinor-helicity basis
can be converted to the massive twistor basis:
$\langle \lambda_2\rambda_2 | S | \lambda_1\rambda_1 \rangle$ $\to$ $\langle \bZ_2 | S | Z_1 \rangle$.
To our best knowledge,
such half-Fourier transforms
have been actively pursued only
for massless legs
\cite{witten2004perturbative,arkani2010s-matrix,Guevara2021reconstructing}.

\paragraph{What Defines Spinning Black Holes}%
Now we ask the following question:
What is the simplest S-matrix in massive twistor space?
For the case of massive-massive-massless three-particle scattering with equal masses,
the on-shell kinematics exhibit two \textit{extremal} cases
that freeze
either of the left-handed or right-handed massive spinor-helicity variables.
Let us call these \textit{zig} and \textit{zag} kinematics
and denote them respectively as white and black blobs
as shown in Fig.\,\ref{fig:onshell},
analogously to the massless on-shell diagram notation
in \rrcite{hodges1980twistor,hodges1990feynman,hodges2005a,hodges2005b,arkani2010s-matrix,arkani2012positivegrassmannian,Atiyah:2017erd}.
Then momentum conservation
fully refines
their delta function supports
as
\begin{align}
    \label{eq:zkin}
\begin{split}
    \text{zig}:\quad
    \del{
        \lambda_1
    }{
        \lambda_2
    }
    \,\mem
    \dell{
        \rambda_1 
            {\mem-\mem} 
        \lambda\inv{1} k_3
    }{
        \rambda_2
    }
    \,,\\
    \text{zag}:\quad
    \dell{
        \lambda_1 
            {\mem-\mem}
        k_3\hem \rambda\inv{2}
    }{
        \lambda_2
    }
    \,\mem
    \del{
        \rambda_1
    }{
        \rambda_2
    }
    \,,
\end{split}
\end{align}
where $k_3{}_{\a\da}$ is the massless momentum of the third leg.
We have abbreviated 
contracted indices as
$(\lambda\inv{1} k_3)_{I\da} = (\lambda\inv{1})_I{}^\a\mem k_3{}_{\a\da}$.
These are \textit{complexified} kinematics
\footnote{
    Cf. Penrose's argument in \rrcite{penrose1976nonlinear,penrose1976curvedtwistor}
    that the notion of a ``single-graviton spacetime''
    is inherently complex.
},
so $\rambda_{1,2}$ here
would be
more appropriately
denoted as $\smash{\tilde{\lambda}_{1,2}}$.

Furthermore,
we stipulate a link between chirality and holomorphy:
the \textit{zig} and \textit{zag} kinematics
are realized for
\textit{positive} and \textit{negative} helicities of the massless leg,
respectively.
With this postulate,
the zig and zag kinematics embody
the scattering amplitudes statement of 
an analog of equivalence principle in half-flat backgrounds,
which we dub \textit{spinning equivalence principle} (SEP).

Suppose a spinning object traveling in a self-dual background spacetime.
What would be its simplest motion?
By definition, the left-handed spinor bundle is flat,
which one can locally trivialize.
If the gravitational forces are to
arise uniquely from this geometry,
the object's left-handed Lorentz frame would remain constant
in local laboratories.
This deduces the SEP:
the left-handed (right-handed) frame
of a spinning object
is covariantly constant
under dynamical evolution in self-dual (anti-self-dual) backgrounds.

Of course, in a modern understanding,
the equivalence principle is rather a statement about the minimality of a test object,
which can be broken by non-minimal couplings.
For example,
a scalar particle endowed with nonzero non-minimal Wilson coefficients
will not follow the geodesic trajectory
as desired by the equivalence principle.
Similarly, the SEP
rather proposes
the most ideal behavior which a spinning object may exhibit.

Remarkably, we find that such an ideal behavior is exhibited by \textit{black holes}.
To see this,
consider the massive half-Fourier transform of the zig kinematics to the twistor \textit{coherent state} basis
\footnote{
    The punchline here parallels that of \rcite{arkani2010s-matrix}.
    Also, a hidden assumption will be that the in and out states for the black hole are given in the twistor coherent states.
}:
\begin{align}
\begin{split}
    \label{eq:half+1}
    &
    \exp\hnem\Big({
        -i\hem\bmu_2
        \lambda_1
    }\Big)
    \,
    \exp\hnem\Big(\mem{
        i\hem(\lambda\inv{1}k_3 \hhnem+\nem\rambda_2)
        \mu_1
    }\Big)
    \\
    &
    =\,
    \BraKet{\bZ_2}{Z_1}
    \,
    \exp\hnem\Big(\mem{
        i\mem k_3\mem\mu_1 \lambda\inv{1}
    }\mem\Big)
    \,.
\end{split}
\end{align}
By the massive incidence relation in \eqref{eq:incidence},
the exponential factor computes to
$e^{ik_3z_1} {\:=\:} e^{ik_3x_1} e^{-k_3y_1}$,
where
the factor $e^{-k_3y_1}$
\footnote{
    Here,
    the minus sign
    is due to our conventions $\ve_{0123} = +1$
    and $S^{\m\n} = \ve^{\m\n\r\s} y_\r\hhem p_\s$:
    $y^\m = -a^\m$.
}
for receiving positive-helicity force carriers
is known to implement
the Newman-Janis shift property
at the amplitudes level
\cite{aho2020,chkl2019,Guevara:2018wpp,Guevara:2019fsj}.
Here, it is crucial that
the ``imaginary impact parameter'' $y_1^\m$ describes spin
due to the very (re)interpretation of 
complex coordinates in twistor theory
as spinspacetime coordinates,
as is deduced in \eqref{eq:incidence}.
Similarly,
half-Fourier transforming the zag kinematics
yields the factor 
$e^{ik_3\bz_1} {\:=\:} e^{ik_3x_1} e^{k_3y_1}$,
encoding the Newman-Janis shift for receiving negative-helicity force carriers.

In sum,
spinning black holes
are characterized by
the extremal complexified on-shell kinematics in Fig.\,\ref{fig:onshell},
which encode the SEP.
Notably, we could have also assumed
any multiplicity of massless quanta,
building up arbitrary half-flat backgrounds.
This derives the same-helicity spin exponentiation \cite{Johansson:2019dnu,Lazopoulos:2021mna,Aoude:2020onz}
at all multiplicities
via the half-Fourier transform.
Eventually, this case would arise from a gluing of the same-helicity on-shell diagrams.
It will be interesting if a smart gluing is viable for mixed helicities as well,
given the spurious pole problem 
\cite{ahh2017,chkl2019,aoude2022searching}.

A few remarks are in order.
Firstly,
in the above derivation of Newman-Janis shift,
the off-diagonal commutator structure
of the massive twistor space
(as the structure of half-Fourier transforms),
namely the K\"ahler geometry,
has played an important role.
To make this point more evident,
consider the following identity:
\begin{align}
\begin{split}
    \label{eq:expl}
    \Bra{\lambda_2\rambda_2}
        \exp\hnem\Big(\mem{
            i\mem 
            k_3 \mu \lambda^{-1}
        }\mem\Big)
    =
    \Bra{\lambda_2,\rambda_2 {\,+\mem} \lambda\inv{1}k_3}
    \,.
\end{split}
\end{align}
Overlapping with $|\lambda_1\rambda_1\rangle$
reproduces \eqref{eq:half+1}
as the matrix element of $e^{ik_3 z}$ in the twistor basis.
Crucially, the \textit{holomorphic} operator $e^{ik_3 z}$
shifts the \textit{anti-holomorphic} variable $\rambda$,
precisely due to the off-diagonal structure.

Another essential role is played by
the intrinsic association between chirality and holomorphy.
To see this, suppose we knew that the object is a spinning black hole from the beginning.
While sitting at the asymptotic infinity
with an experimentalist mindset,
we ``measure'' the S-matrix to deduce what background this black hole had interacted with.
If the S-matrix is \textit{holomorphic},
then, by the off-diagonal structure,
the \textit{anti-holomorphic} variable
$\rambda$
solely exhibits precession.
But crucially, the \textit{anti-holomorphic} spinor-helicity variable is \textit{right-handed}.
Due to this very endemic link between holomorphy and handedness,
the background must have been
filled with \textit{right-handed}, i.e., self-dual fields.

On a related note,
the exponent in \eqref{eq:expl}
describes the self-dual angular momentum
$J^+ = (z\mwedge p)^+$
times a factor,
providing a twistorial realization of the equations employed in \rcite{Guevara:2018wpp}.

For generic massive spinning objects,
the kinematics is not extremal.
For example,
$\lambda$ and $\rambda$
are shifted symmetrically (cf.~\rcite{aho2020})
when
the object's spin-induced multipole moments \cite{Levi:2015msa} vanish from the quadrupole order
\cite{ambikerr1}.
The complexified kinematics
we have studied
is sensitive to all multipole moments,
while Newman's explorations \cite{newman1974curiosity,newman1974collection,newman2004maxwell,Newman:1973yu,Newman:2002mk}
focused only on the dipole coupling (gyromagnetic ratio).

Lastly,
we were able to deduce the essential physics
just by focusing on the delta function support,
similarly as in \rcite{arkani2012positivegrassmannian}.
A full-blown theory should not only incorporate the $mx^h$ prefactors \cite{ahh2017}
but also properly formulate
the independent massive little group orbits.
The first-quantized massive twistor particle \cite{ambikerr0,ambikerr1}
could provide a reference,
which concretely reproduces the delta function supports
considered in this paper
\cite{ambikerr1}.

\paragraph{Conclusions}%
In this work, we established
the notion of spinspacetime as a universal consequence of Poincar\'e symmetry.
Its characteristic geometric structures,
which trace back to those of twistor space,
revealed that spinning black holes 
obey a spinning analog of equivalence principle.
The Newman-Janis shift factors 
are effortlessly obtained by
massive half-Fourier transforming
the extremal on-shell kinematics in \fref{fig:onshell}.

This revival of a curious concept
triggers various future directions.
Firstly,
the theory of bulk spinspacetime will be developed
\cite{sodual,sst-earth}.
Secondly,
applications to post-Minkowskian gravity
will be pursued.
For instance, 
the supplemental material
suggests
simpler approaches to
the effective potential 
\cite{Chung:2018kqs,Chung:2019duq,Chung:2020rrz}
or Poincar\'e-covariant center coordinates
\cite{newman1974curiosity,newman1974collection,Regge:1974zd,Damour:2000kk,Damour:2007nc,Lee:2023nkx}.
Lastly,
the massive on-shell diagrams
and their gluing
should be investigated,
reminiscently of the massless story
\cite{hodges1980twistor,hodges1990feynman,hodges2005a,hodges2005b,arkani2010s-matrix,arkani2012positivegrassmannian,Atiyah:2017erd,witten2004perturbative}.
It will be exciting if
a complete on-shell framework,
incorporating both massless and massive legs,
can serve as a powerful tool
for both theoretical and practical directions.

\medskip
\paragraph{Acknowledgements}%
    The author would like to thank
        Nima Arkani-Hamed,
        Clifford Cheung,
        Thibault Damour,
        Jung-Wook Kim,
        Sangmin Lee,
        Keefe Mitman,
        Alexander Ochirov,
    and
        Julio Parra-Martinez
    for discussions or comments.
    This material is based upon work supported by the U.S. Department of Energy, Office of Science, Office of High Energy Physics, under Award Number DE-SC0011632.
    \JHK is also supported 
    by Ilju Academy and Culture Foundation.

\input{supp.tex}

\bibliography{references.bib}

\end{document}

%% file: supp.tex
\newpage
\onecolumngrid

\appendix

\section*{Bootstrapping spinspacetime}
\label{app:bootstrap}

In the main text,
we established the \textit{existence} of spinspacetime from Poincar\'e charges.
Here, we establish its \textit{uniqueness}.
Specifically,
we show the inevitability of complex-valued coordinates
for achieving both commutativity and Poincar\'e covariance,
in the format of a bootstrap.

The input of the bootstrap is the following.
First, suppose Poincar\'e charges
$J^{\m\n}$ and $p_\m$
with the dilatation charge
$D$ defining the notion of mass dimension,
with $p^2 < 0$.
\eqrefs{eq:pa}{eq:D|Jp}
have described their commutator algebra.
Second, the following commutation relations
are taken
as the definition of ``spacetime coordinates'' $q^\m$:
\begin{subequations}
    \label{eq:x-axioms}
\begin{align}
    \label{eq:Jcond}
    [ q^\m , J^{\r\s} ]
    &= 
    i\mem
    (-2\eta^{\m[\r} \delta^{\s]}{}_\n)\mem
    q^\n
    \,,\\
    \label{eq:Pcond}
    [ q^\m , p_\r ]
    &= 
    i\mem
    \delta^\m{}_\r
    \,,\\
    \label{eq:Dcond}
    [ q^\m , D ]
    &= -
    i\mem 
    q^\m
    \,.
\end{align}
\end{subequations}
In other words,
a set of four variables $q^\m$
defines Poincar\'e-covariant spacetime coordinates iff it satisfies Eqs.\,\oldeqref{eq:Jcond}, \oldeqref{eq:Pcond}, and \oldeqref{eq:Dcond}.
These
specify the desired transformation behavior of $q^\m$ under the Poincar\'e action
and also its mass dimension $-1$.
Third, the commutativity condition for such coordinates is
\begin{align}
    \label{eq:commutecond}
    [ q^\m , q^\n ] = 0
    \,.
\end{align}
The goal of this bootstrap
is to find $q^\m$
in terms of $p_\m$, $J^{\m\n}$, $D$
when all of these three conditions are true.

To begin with,
we impose \eqref{eq:x-axioms},
namely the axioms for spacetime coordinates.
First of all,
\eqref{eq:Jcond}
implies that
$q^\m$ cannot depend on any
constant tensors
carrying one or more indices,
as they break Lorentz symmetry as reference structures.
By considering \eqref{eq:Dcond} as well,
it then follows that
$q^\m$ can be a linear combination of
\begin{align}
    \label{eq:tower1}
    p^\m \nem/p^2
    \,,\quad
    \hat{x}^\m
    := J^{\m\n} p_\n /p^2
    \,,\quad
    y^\m
    := -{*}J^{\m\n} p_\n /p^2
    \,,\quad
    J^\m{}_\n\mem \hat{x}^\n
    \,,\quad
    J^\m{}_\r\mem J^\r{}_\n\mem \hat{x}^\n
    \,,\quad
    \cdots
    \,,
\end{align}
with mass dimension zero coefficients such as
\begin{align}
    \label{eq:tower2}
    D
    \,,\quad
    J_{\m\n}\mem J^{\m\n}
    \,,\quad
    p_\m\mem J^\m{}_\r\hem J^\r{}_\n\mem p^\n \nem/ p^2
    \,,\quad
    \cdots
    \,.
\end{align}
A linearly independent basis
may be further set up
by employing the identity
$({*A})^{\m\r}\hem ({*B})_{\r\n}
= -\frac{1}{2}\mem \delta^\m{}_\n\mem
(A_{\r\s} B^{\r\s})
- A^{\m\r} B_{\r\n}$
for antisymmetric tensors $A^{\m\n}$, $B^{\m\n}$.

In \eqref{eq:tower1},
the terms 
are enumerated in the powers of $J$.
Upon imposing the condition in \eqref{eq:Pcond},
however,
this tower gets truncated at the linear order in $J$:
$p^\m \nem/p^2$, $\hat{x}^\m$, $y^\m$.
This is because
taking commutator with $p$
can eliminate at most one $J$,
while
the right-hand side of \eqref{eq:Pcond}
contains no $J$.
Similarly, the tower in \eqref{eq:tower2}
also gets truncated 
so that all the coefficients
should be sole functions of $D$:
there is no scalar combination
that involves only one $J$,
due to its antisymmetry.
As a result of these considerations,
$q^\m$ must take the following form:
\begin{align}
    q^\m
    = \alpha_0(D)\mem p^\m \nem/p^2
    + \alpha_1(D)\mem \hat{x}^\m
    + \alpha_2(D)\mem y^\m
    \,.
\end{align}
Then imposing \eqref{eq:Pcond}
yields
$
\delta^\m{}_\r
= 
    \alpha_1(D)\mem (\delta^\m{}_\r - p^\m p_\r /p^2)
    -\alpha_0'(D)\mem p^\m p_\r /p^2
    -\alpha_1'(D)\mem p_\r\mem \hat{x}^\m
    -\alpha_2'(D)\mem p_\r\mem y^\m
$,
from which it follows that
$\alpha_0'(D) = -1$, $\alpha_1(D) = 1$, and $\alpha_2'(D) = 0$.
In turn, the form of $q^\m$ gets constrained to
\begin{align}
\begin{split}
    q^\m
    \,=\, (c_0{\,-\,}D)\mem p^\m \nem/p^2
    + \hat{x}^\m
    + c_2\mem y^\m
    \,=\, c_0\mem p^\m \nem/p^2
    + x^\m
    + c_2\mem y^\m
    \,,
\end{split}
\end{align}
where $c_0$ and $c_2$ are constants.
Here,
$x^\m = \hat{x}^\m - D\mem p^\m \nem/p^2$ are the spacetime coordinates
used in the main text: \eqref{eq:xdef}.

Finally, 
we impose the commutativity condition in \eqref{eq:qq}.
The commutator $[q^\m,q^\n]$ evaluates to
$
    [ x^\m , x^\n ]
    + 2{c_2} [ x^{[\m} , y^{\n]} ]
    + {c_2}^2 [ y^\m , y^\n ]
$,
while
straightforward computation using 
the definition of $\hat{x}^\m$, $y^\m$
with
\eqrefs{eq:pa}{eq:D|Jp}
derives
$[ x^\m , x^\n ] = -i\mem S^{\m\n} \nem/p^2$
and 
$[ x^{\m} , y^{\n} ] = -2i\mem y^{(\m} p^{\n)} \nem/p^2$.
Consequently, it is found that
\begin{align}
    \label{eq:qq}
    [ q^\m , q^\n ]
    = -\frac{i\mem(1{\,+\,}{c_2}^2)}{p^2}\mem S^{\m\n}
    \,,
\end{align}
from which it follows that $c_2 = \pm\mem i$:
\begin{align}
\begin{split}
    q^\m
    &= c_0\mem p^\m \nem/p^2
    + 
    (x^\m
    \pm i\mem y^\m)
    \,.
\end{split}
\end{align}

Therefore,
the only possible \textit{Poincar\'e-covariant and commutative}
coordinates
are the holomorphic or anti-holomorphic spinspacetime coordinates 
$x^\m \pm iy^\m$
given in the main text,
up to a constant shift along the direction longitudinal to the momentum
with a \textit{c-number} coefficient.

Said in another way,
there does not exist
a Poincar\'e-invariant notion of \textit{real commutative} spacetime.
For any real $c_2$, the right-hand side of \eqref{eq:qq} is nonzero.
To make it vanish,
the use of complex numbers is unavoidable.

The reader might wonder
how exactly
a set of 
real-valued 
commutative coordinates
could be achieved
by relaxing some of our bootstrap conditions.
For the sake of concreteness,
let us
explicitly
spell out
one such possibility:
\begin{align}
\begin{split}
    \label{eq:rmx}
    \rmx^\m
    := 
    \frac{1}{\P_\k\hem p^\k}\mem
    \bigg(\hem{
        \delta^\m{}_\r
        - \frac{p^\m p_\r}{p^2}
    }\bigg)\mem
    J^{\r\n} \P_\n
    - \frac{p^\m}{p^2}\mem D
    \,,\quad
    \P_\m
    := p_\m + (-p^2)^{1/2}\mem l_\m
    \,,
\end{split}
\end{align}
where $l_\m$ is constant.
It is straightforward to show that
$q^\m = \rmx^\m$
satisfies \eqrefs{eq:Pcond}{eq:Dcond}.
A lengthy calculation further shows that
$
    [ \rmx^\m , \rmx^\n ]
    = 
        i\mem (1{\,+\,}l^2)\mem S^{\m\n}
        /
            (\hem{
                \P \cdot\hnem l
                - (-p^2)^{1/2} (1{\,+\,}l^2)
            })^2
$,
so \eqref{eq:commutecond} is satisfied iff $l_\m$ is chosen to be unit timelike:
$l^2 {\:=\:} {-1}$.
Of course,
\eqref{eq:Jcond}
is violated since \smash{$l_\m$} is a Lorentz-violating
reference structure:
$
    \smash{[ \rmx^\m , J^{\r\s} ]}
    =
    i\mem
    (-2\eta^{\m[\r} \delta^{\s]}{}_\n)\mem
    \rmx^\n
    + (i/\P\cdot p)\mem \rmN^{\m|\r\s}
$,
$
    \rmN^{\m|\r\s}
    =
        2S^{\m[\r} (\mathcal{P}{\mem-\,}p)^{\s]}
    + 
        (x{\,-\,}\rmx)^\m\mem 
        (\mathcal{P}\kern-0.035em \mwedge p)^{\r\s}
$.
Yet, the $\mathrm{SO}(3)$ covariance under
the rotations normal to $l^\m$
is preserved,
as \smash{$l_\r\mem {*}\rmN^{\m|\r\s} = \frac{1}{2}\mem l^\r\mem \ve_{\r\s\k\l}\mem \rmN^{\m|\k\l} = 0$}.
Also, the total angular momentum
is split as
$
    J^{\m\n}
    = 2\rmx^{[\m}p^{\n]}
    + \rmS^{\m\n}
$ with $
    \rmS^{\m\n}
    =
    J^{\m\n}
    - 
    (
        p^\m \P_\r J^{\r\n}
        +
        p^\n \P_\r J^{\m\r}
    )
    /p^\k\P_\k
$,
so $\P_\m\mem \rmS^{\m\n} = 0$.
This is exactly
the Pryce-Newton-Wigner (canonical)
spin supplementary condition
\cite{pryce1948mass,Newton:1949cq,steinhoff2015spin}.
The upshot here is that reality engages in a trade-off relationship with Lorentz covariance, 
given translation covariance and commutativity.

\section*{Spinspacetime for Black Hole Binaries}
\label{app:center}

Here, we sketch some applications of spinspacetime.
We hope to release more details in future works.

First,
consider the gluing of the on-shell diagrams in \fref{fig:onshell}
by the massless line.
When all the massive legs
are half-Fourier transformed
to the twistor coherent state basis,
one finds factors of $e^{ik(z_1-\bz_2)}$ or $e^{ik(\bz_1-z_2)}$
for $k$ the massless momentum,
where $z_{1,2},\bz_{1,2}$
are the spinspacetime coordinates associated with the two massive lines.
By incorporating the $mx^h$ prefactors \cite{ahh2017}
and using a well-known identity \cite{aho2020,Guevara:2017csg},
the leading-order post-Minkowskian (PM) effective potential
for Kerr black hole binaries is found
in the zero-momentum frame
as
\begin{align}
    \includegraphics[scale=0.5,valign=c]{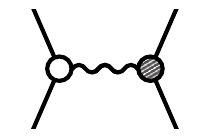}
    {\kern-0.5em}
    \,+\,
    {\kern-0.5em}
    \includegraphics[scale=0.5,valign=c]{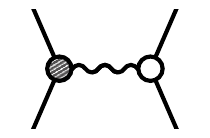}
    \,\,\rightsquigarrow\,\,
    \,\,\,\mem
    H^{(1)}
    \,\sim\,\,\hem
    -\frac{G{m_1}^2{m_2}^2}{2E_1E_2}\mem
    \bigg(\,{
        \frac{
            e^{+2\varphi}
        }{\big|\vec{z}_1{\,-\,}\vec{\bz}_2\big|}
        \,+\,
        \frac{
            e^{-2\varphi}
        }{\big|\vec{\bz}_1{\,-\,}\vec{z}_2\big|}
    }\,\bigg)
    \,,
\end{align}
where $\varphi$ denotes the relative rapidity.
This expression can be brought to the known one
\cite{Chung:2018kqs,Chung:2019duq,Chung:2020rrz}
by changing the spin supplementary condition
from covariant to canonical.

Second, the asymptotic spinspacetime coordinates can be thought of as the complex center of mass
\cite{newman1974curiosity,newman1974collection}.
The simplest example is a non-interacting binary system of spinning bodies in flat spacetime.
The conformal charges are additive.
Remarkably, \eqref{eq:z=Gp} then offers a fascinatingly simple derivation of
Poincar\'e-covariant center coordinates:
\begin{align}
    \label{eq:zcen}
    z^{\da\a}
    \mem=\mem
    (z_1)^{\da\b}\hem (p_1p^{-1})_\b{}^\a
    +
    (z_2)^{\da\b}\hem (p_2p^{-1})_\b{}^\a
    \,,
\end{align}
where $p_{\a\da} = (p_1 {\,+\,} p_2)_{\a\da}$
is the total momentum.
In particular, the real part of \eqref{eq:zcen} reads
\begin{align}
\begin{split}
    \label{eq:xcen}
\begin{aligned}
    x^\m
    \mem=\mem
    (e_1 x_1 + e_2 x_2)^\m
    &
    -
    (p_1 \mwedge p_2)^\m{}_\n
    \hem(x_1 {\mem-\,} x_2)^\n
    \nem/p^2
    +
    {*}(p_1 \mwedge p_2)^\m{}_\n
    \hem(y_1 {\mem-\,} y_2)^\n
    \nem/p^2
    \,.
\end{aligned}
\end{split}
\end{align}
The first term describes the average position
from the energy weights in the zero-momentum frame,
$e_{1,2} = p_{1,2} \mdot (p^{-1}) = (-p \mdot p_{1,2}) /(-p^2)$.
The second term
adds a
transversal correction,
enforcing additivity of the dilatation charges.
This promotes
the formula 
to
a map between worldlines
\smash{$x_{1,2}^\m \sim x_{1,2}^\m + \varkappa_{1,2}\mem p_{1,2}^\m$},
$x^\m \sim x^\m + \varkappa\mem p^\m$.
The third term is a spin-dependent correction,
which can be attributed to the electric-magnetic duality of Newman \cite{newman1974curiosity}.
Indeed,
the imaginary part of \eqref{eq:zcen}
offers
an intriguing notion of
``spin center,''
exhibiting also an electric-magnetic duality 
with \eqref{eq:xcen}:
\begin{align}
\begin{split}
    \label{eq:ycen}
\begin{aligned}
    y^\m
    \mem=\mem
    (e_1 y_1 + e_2 y_2)^\m
    &
    -
    (p_1 \mwedge p_2)^\m{}_\n
    \hem(y_1 {\mem-\,} y_2)^\n
    \nem/p^2
    -
    {*}(p_1 \mwedge p_2)^\m{}_\n
    \hem(x_1 {\mem-\,} x_2)^\n
    \nem/p^2
    \,.
\end{aligned}
\end{split}
\end{align}
Note that
$y^\m$ does not simply vanish by sending the individual spins $y_1^\m$, $y_2^\m$ to zero,
as
the binary system can still have a nonzero ``spin'' angular momentum
when effectively considered as a single object.

When there is a weak interaction between the two bodies,
perturbatively expanding the conformal charges
yields
the PM corrections to the zeroth-order formula in 
\eqref{eq:zcen}:
$ 
    z {\mem\,+\mem} \mathit{\Delta}z
    = (
        zp
        {\,-\mem} \mathit{\Delta}G
    )\mem
    (p {\mem\,+\mem} \mathit{\Delta}p)^{-1}
$,
implying
$z^{(1)} = (- G^{(1)} p^{-1}) - z\mem (p^{(1)} p^{-1})$
at the leading order.
But crucially,
the $\mathrm{ISO}(3)$ symmetry
in the zero-momentum frame
mandates that
the conformal charges are deformed in a particular form:
$\mathit{\Delta} G^\da{}_\wrap{\db} = - \mathit{\Delta} X^{\da\a} p_\wrap{\a\db}$,
$\mathit{\Delta} p_{\a\da} = (H/E{\,-\,}1)\mem p_{\a\da}$,
where 
$\mathit{\Delta} X^\m$ is real.
Here, \smash{$E = (-p^2)^{1/2}$} is the 0PM Hamiltonian.
Imposing the full Poincar\'e algebra
then gives
\begin{align}
    (G^{(1)})^\da{}_\db
    \,=\, 
    (H^{(1)}\nem/E)\mem 
        ({ (e_2 x_1 {\,+\,} e_1 x_2) \wedge p}\mem)^\da{}_\db
    \qiq
    z^{(1)}{}^\m
    \,=\,
    (H^{(1)}\nem/E)\mem 
    (
        \delta^\m{}_\n {\mem+\,} p^\m p_\n / E^2
    )\mem
    (
        e_2 x_1 {\,+\,} e_1 x_2
        - z
    )^\n
    \,,
\end{align}
which reproduces the results in \rcite{Lee:2023nkx}
upon changing the spin supplementary condition to covariant to canonical.
Notably,
the general lesson
seems that
the covariant complex spinspacetime coordinates
could offer the best shortcut
for deriving various formulae
even when they are supposed to be obtained in non-covariant spin gauges.

\newpage
\twocolumngrid